\documentstyle[amssymb,preprint,aps]{revtex}

\begin{document}
\draft
\title{Drift of a polymer chain in disordered media}
\author{Semjon Stepanow and Michael Schulz}
\address{Martin-Luther-Universit\"{a}t Halle-Wittenberg, Fachbereich Physik, D-06099\\
Halle, Germany}
\date{\today }
\maketitle

\begin{abstract}
We consider the drift of a polymer chain in a disordered medium, which is
caused by a constant force applied to the one end of the polymer, under
neglecting the thermal fluctuations. In the lowest order of the perturbation
theory we have computed the transversal fluctuations of the centre of mass
of the polymer, the transversal and the longitudinal size of the polymer,
and the average velocity of the polymer. The corrections to the quantities
under consideration, which are due to the interplay between the motion and
the quenched forces, are controlled by the driving force and the degree of
polymerization. The transversal fluctuations of the Brownian particle and of
the centre of mass of the polymer are obtained to be diffusive. The
transversal fluctuations studied in the present Letter may also be of
relevance for the related problem of the drift of a directed polymer in
disordered media and its applications. 
\end{abstract}


The behaviour of a polymer chain in disordered media is of interest for
different applications such as gel electrophoresis of biopolymers (see \cite
{noolandi87}-\cite{rotstein/lodge92} and references therein). It is also
expected that the study of the behavior of a polymer in a disordered medium
may be useful for a better understanding such a fundamental problem as
reptation \cite{degennes71}-\cite{doi/edwards86}. Despite the large interest
in this problem in the last decade \cite{edwards/muthu88}-\cite
{baumg/muthu87} no complete quantitative understanding of the behaviour of
the polymer chain has been achieved so far. The reason is the quenched
nature of the disorder, which after the average results in an effective
attraction between the polymer segments and is responsible that the problem
under consideration becomes a strong coupling problem. The standard
theoretical tools such as perturbation theories and also the renormalization
group method break down in a number of interesting cases for these kind of
problems due to the runaway of the effective coupling constant. Thus, the
polymer in a disordered media is characterized by a nontrivial interplay
between the thermal effects and the disorder, which so far is beyond of a
quantitative analytical treatment. The interplay between disorder and
thermal effects is also of great importance for different kind of creep
phenomena \cite{ioffe/vinokur86}-\cite{ledoussal98} (and references therein).

In the present Letter we will present the results of the study of the
simpler problem of the drift of a polymer chain in disordered media under
the action of a constant force applied to one end of the polymer by
neglecting the thermal effects. The motivation of the present problem is the
electrophoresis of biopolymer molecules in porous media. The drift of a
polymer in disordered media can be considered as a further example of the
driven dynamics of manifolds (flux lines, interfaces, etc.) in disordered
media, where a considerable progress was achieved in last years \cite{nstl92}%
. One of the interesting aspects of the driven dynamics of the polymer chain
in a disordered medium is the generation of the transversal fluctuations,
which is due to the interplay between the motion and the quenched random
forces. This effect, which to our knowledge have not been studied
previously, appears already in the case of a Brownian particle (at zero
temperature). We will argue in this Letter that the transversal fluctuations
of the centre of mass of the polymer chain (and of the Brownian particle)
generated due to the interplay between the disorder and motion are
diffusive. We expect that such transversal fluctuations may be of interest
for directed manifolds (flux lines, crack propagation, etc.).

\bigskip We start the consideration with the Langevin equation for a
discrete polymer chain \cite{doi/edwards86} 
\begin{equation}
f\frac{\partial r_{i}}{\partial t}=\varkappa
(r_{i+1}+r_{i-1}-2r_{i})+F_{i}(r_{i})+\eta _{i},  \label{dp1}
\end{equation}
where $r_{i}^{\alpha }$ give the positions of monomers ($i$ = $1,...,N$) in $%
d$-dimensional space ($\alpha $ = $1,...,d$), $f$ is the monomer friction
coefficient. The elastic forces are associated with the potential $U=\frac{%
\varkappa }{2}\sum_{i=1}^{N}(r_{i}-r_{i-1})^{2}$. Notice that the elasticity
constant $\varkappa $ is given by $\varkappa =dkT/l^{2}$, $l$ is the
statistical segment length, and $N=L/l$ is the number of monomers. In the
following we will consider the drift of the polymer chain at zero
temperature, so that we will consider $\varkappa $ as a constant, which is
temperature independent. The force $F_{i}(r)=F\delta _{i1}+g(r)$ with $F$
being the constant driving force acting on the $1st$ segment of the polymer,
and $g(r)$ is the random quenched force. We assume that the latter is
Gaussian distributed with the zero mean and the correlator 
\begin{equation}
\left[ g^{\alpha }(r)g^{\beta }(r^{\prime })\right] =C^{\alpha \beta
}(r-r^{\prime }).  \label{dp2}
\end{equation}
The Fourier transform of $C^{\alpha \beta }(r)$, $C^{\alpha \beta }(q)$, is
assumed to decay exponentially for large $q$ with a width $a$. In particular 
$C^{\alpha \alpha }(q)$ has the shape $C^{\alpha \alpha }(q)=\Delta \exp
(-a^{2}q^{2})$ with $\alpha =x,y,z$ in $3d$. The thermal forces are Gaussian
distributed with the correlator given by 
\begin{equation}
<\eta _{i}^{\alpha }(t)\eta _{j}^{\beta }(t^{\prime })>\,=2\delta ^{\alpha
\beta }\delta _{ij}fkT\delta (t-t^{\prime }).  \label{dp3}
\end{equation}
The Langevin equation (\ref{dp1}) can be formulated in a standard way by
using the path integrals, which in the limit of a continuous chain \cite
{doi/edwards86} results in the following expression for the transition
probability density for a polymer configuration $r(s,t)$   
\begin{equation}
P(r(s),t;r^{0}(s),t_{0})=\int_{r(s,t_{0})=r^{0}(s)}^{r(s,t)=r(s)}Dr(s,t)\int
Dp(s,t)\exp (-S),  \label{dp4}
\end{equation}
where the action for the continuous polymer chain reads 
\begin{equation}
S=D\mu ^{2}\int_{0}^{L}ds\int_{t_{0}}^{t}dt^{\prime }p(s,t^{\prime
})^{2}-i\int_{0}^{L}ds\int_{t_{0}}^{t}dt^{\prime }p(s,t^{\prime })(\frac{%
\partial r(s,t^{\prime })}{\partial t^{\prime }}-\gamma \mu \nabla
_{s}^{2}r(s,t^{\prime })-\mu g_{s}(r(s,t^{\prime }))  \label{dp5}
\end{equation}
with $D=fkT/l$, $\mu ^{-1}=f/l$, $\gamma =\varkappa l$, and $l$ being the
statistical segment length. The generating functional of the non interacting
polymer, $Z_{0}(j,\widetilde{j})$ $=\int_{r(s,t_{0})=r^{0}(s)}Dr(s,t)\int
Dp(s,t)\exp (-S+i\int_{0}^{L}ds\int_{t_{0}}^{t}dt^{\prime }(r(s,t^{\prime
})j(s,t^{\prime })+p(s,t^{\prime })\widetilde{j}(s,t^{\prime }))$, is given
by 
\begin{eqnarray}
Z_{0}(j,\widetilde{j}) &=&\exp
(-i\int_{t_{0}}^{t}dt_{1}%
\int_{t_{0}}^{t}dt_{2}j_{k}(t_{1})G_{k}^{0}(t_{1}-t_{2})\widetilde{j}%
_{k}(t_{2})+i\int_{t_{0}}^{t}dt_{1}j_{k}(t_{1})G_{k}^{0}(t_{1}-t_{2})\xi
_{k}^{0}+  \nonumber \\
&&i\int_{t_{0}}^{t}dt_{1}%
\int_{t_{0}}^{t}dt_{2}j_{k}(t_{1})G_{k}^{0}(t_{1}-t_{2})\mu g_{k}-  \nonumber
\\
&&\frac{D\mu ^{2}}{2\gamma \mu k^{2}}\int_{t_{0}}^{t}dt_{1}%
\int_{t_{0}}^{t}dt_{2}j_{k}(t_{1})(G_{k}^{0}(t_{1}-t_{2})+G_{k}^{0}(t_{2}-t_{1})-G_{k}^{0}(t_{1}+t_{2}-2t_{0}))j_{k}(t_{2}),
\label{dp6}
\end{eqnarray}
where the Fourier transform of a quantity $a(s)$ is defined as follows $%
a(s)=\sum_{k=0}^{\infty }Q_{sk}a_{k}$ with $Q_{sk}=\sqrt{2/L}\cos (\pi sk/L)$%
, and $Q_{s0}=1/\sqrt{L}$, $G_{k}^{0}(t)=\exp (-\gamma \mu k^{2}t)\theta (t)$
with $\theta (t)$ being the Heaviside function, and $\xi _{k}^{0}$ is the
Fourier transform of $r(s,t=0)$. The last term in the exponential of Eq.(\ref
{dp6}) disappears for zero temperatures. The generating functional $Z(j,%
\widetilde{j}=0)$ at zero temperature can be written in the symbolic form as 
\begin{eqnarray}
Z(j) &=&\exp
(\int_{0}^{t}dt_{1}\int_{0}^{t}dt_{2}\int_{0}^{L}ds_{1}\int_{0}^{L}ds_{2}%
\int_{q}j_{k}^{\alpha }(t_{1})G_{k}^{0}(t-t_{1})Q_{s_{1}k}C^{\alpha \beta
}(q)Q_{s_{2}k}G_{k}^{0}(t-t_{2})j_{k}^{\beta }(t_{2})+  \nonumber \\
&&i\int_{t_{0}}^{t}dt_{1}%
\int_{t_{0}}^{t}dt_{2}j_{k}(t_{1})G_{k}^{0}(t_{1}-t_{2})\mu g_{k}),
\label{dp7}
\end{eqnarray}
where the source $j_{k}(t)$ in Eq.(\ref{dp7}) has to be modified in each
order of the perturbation expansion as follows: $j_{k}(t)\rightarrow
j_{k}(t)+qQ_{s_{1}k}\delta (t-t_{1})-qQ_{s_{2}k}\delta (t-t_{2})$, where $q$%
, $s_{1}$, $s_{2}$, $t_{1}$, and $t_{2}$ are variables associated with the
disorder correlator, in the $1st$ order, $j_{k}(t)\rightarrow
j_{k}(t)+q_{1}Q_{s_{1}k}\delta (t-t_{1})-q_{1}Q_{s_{2}k}\delta
(t-t_{2})+q_{2}Q_{s_{3}k}\delta (t-t_{3})-q_{2}Q_{s_{4}k}\delta (t-t_{4})$
in the $2nd$ order and so on.

We now will study the behaviour of the ideal polymer chain under the action
of the force acting at one end of the polymer under neglecting thermal
fluctuations. Due to the modification of $j_{k}(t)$ the last term in (\ref
{dp7}) gives rise (in the steady state limit) to the sum $%
\sum\limits_{k=1}^{\infty }(Q_{s_{1}k}-Q_{s_{2}k})/(\pi k/L)^{2}=\sqrt{L/2}%
(s_{2}-s_{1})(1-(s_{1}+s_{2})/2L)$. Both the transversal fluctuations of the
centre of mass and the transversal size of the polymer can be expressed by
the quantity 
\begin{equation}
\left[ \xi _{k_{1}}^{\alpha }(t)\xi _{k_{2}}^{\beta }(t)\right] =\frac{1}{%
i^{2}}\delta ^{2}Z(j)/\delta j_{k_{1}}^{\alpha }(t)\delta j_{k_{2}}^{\beta
}(t).  \label{dp8}
\end{equation}
The transversal size of the polymer is obtained from Eq.(\ref{dp8}) with $%
k_{1}$ and $k_{2}$ nonzero and $\alpha $, $\beta \neq z$. The analytical
expression associated with Eq.(\ref{dp8}) reads 
\begin{eqnarray}
&&\mu
^{2}l^{-2}\int_{0}^{L}ds_{1}\int_{0}^{L}ds_{2}\int_{0}^{t}dt_{2}%
\int_{0}^{t_{2}}dt_{1}\int_{q}C^{\alpha \beta
}(q)Q_{s_{1}k_{1}}Q_{s_{2}k_{2}}G_{k_{2}}^{0}(t-t_{2})G_{k_{1}}^{0}(t-t_{1})
\nonumber \\
&&\exp (iq_{z}g_{k}(t_{1}-t_{2})+iq^{\prime
}(s_{2}-s_{1})(1-(s_{1}+s_{2})/2L)),  \label{dp9}
\end{eqnarray}
where $g_{k}=\mu F/L$ and $q^{\prime }=q_{z}F/(\sqrt{2}\gamma )$. Due to the
fact that we are interested in large $t$, we have neglected the time
dependent part associated with the second term in the exponential of Eq.(\ref
{dp9}). Integrations over times in (\ref{dp9}) can be performed
straightforwardly. The integrals over $s_{1}$ and $s_{2}$ in (\ref{dp9}) can
be also computed analytically. It appears that the main contribution in the
latter originates from $k_{1}=k_{2}$ (the term with $k_{1}\neq k_{2}$
vanishes for large $F$), so that we obtain 
\begin{eqnarray}
A(k,q) &=&\int_{0}^{L}ds_{1}\int_{0}^{L}ds_{2}\cos (ks_{1})\cos (ks_{2})\exp
(iq(s_{2}-s_{1})(1-(s_{1}+s_{2})/2L))=  \nonumber \\
&&(\pi L/4q)((C(\sqrt{L/\pi q}(q+k))+C(\sqrt{L/\pi q}(q-k))^{2}+  \nonumber
\\
&&(S(\sqrt{L/\pi q}(q+k))+S(\sqrt{L/\pi q}(q-k))^{2}),  \label{dp10}
\end{eqnarray}
where $C(x)$ and $S(x)$ are Fresnel functions. In the limit of large $L$ the
function $A(k,q)$ tends to the function $(\pi L/2q)(\theta (\mid q\mid -\mid
\,k\mid )$. After performing the integrations over $t_{i}$ and $s_{i}$ ($%
i=1,2$) in Eq.(\ref{dp8}) and using the property of $A(k,q)$ for large $L$,
the mean-square transversal size of the polymer chain is obtained to the
first order in disorder strength as 
\begin{equation}
\left[ \Delta r_{tr}^{2}\right] =\frac{(d-1)a\sqrt{2}L^{3}}{4\gamma l^{2}F}%
\pi ^{d-1}\Delta /a^{d}+O(\Delta ^{2}).  \label{dp11}
\end{equation}
The fluctuations of the centre of mass of the polymer chain to the first
order in disorder strength can be studied by using (\ref{dp8}-\ref{dp9})
with $k_{1}=k_{2}=0$. The mean-square transversal displacement of the centre
of mass of the polymer chain is obtained after performing the integrations
over times $t_{1}$, $t_{2}$ and $s_{1}$, $s_{2}$ as 
\begin{equation}
\left[ \Delta r_{c}^{2}(t)\right] =\frac{(d-1)atL\mu }{l^{2}F}\pi
^{d-1}\Delta /a^{d}+O(\Delta ^{2})  \label{dp12}
\end{equation}
Setting $L=l$ in Eq.(\ref{dp12}) yields the mean-square transversal
displacement for a Brownian particle.

We now will derive Eq.(\ref{dp12}) for a Brownian particle in a qualitative
way. Due to the motion in the direction of the force $F$, the particle
experiences the random force in the transversal directions too. The
transversal fluctuations can be estimated according to the Langevin equation
of the particle as $f^{2}x_{tr}^{2}/t^{2}\simeq \delta F^{2}$ with $\delta
F^{2}\simeq a\Delta /a^{d}/l_{F}$ being the square of the typical value of
the random force and $l_{F}=\mu Ft$ being the characteristic length. It
follows from Eq.(\ref{dp12}) and from the above qualitative consideration
that the transversal behaviour of the centre of mass of the polymer chain
and of the Brownian particle is diffusive. The analysis of higher-order
corrections to (\ref{dp12}) shows that they are linear in $t$, so that the
transversal fluctuations of the particle are expected to be diffusive.
Rewriting (\ref{dp12}) in terms of the unperturbed velocity $v_{c,0}$ $=\mu
F/L$ yields that the mean-square transversal displacements for both the
Brownian particle and the polymer chain coincide.

Excluding the time $t$ in (\ref{dp12}) in favor of the arc length $L$ by
using the relation $\mu \gamma t\sim L^{2}$, which is obtained from the
unperturbed response function $\exp (-\gamma \mu (\pi k/L)^{2}t)$ or from
the relation $v_{c}t\simeq \delta r^{z}$ with $v_{c}$ and $\delta r^{z}$
given by Eqs.(\ref{dp16},\ref{dp18})\footnote{%
The latter argument was suggested by one of the referee.} up to order $%
O(\Delta ^{0})$, we find that Eq.(\ref{dp12}) is compatible with Eq.(\ref
{dp11}). To our knowledge the transversal fluctuations given by (\ref{dp11}-%
\ref{dp12}) have not been studied before. Notice that the temperature is
zero and the transversal fluctuations are due to the interplay between the
motion and the {\it frozen in} random forces.

To study the average velocity of the polymer we will consider the quantity 
\begin{equation}
\left[ \xi _{k}^{z}(t)\right] =i^{-1}\delta Z(j)/\delta j_{k}^{z}(t)
\label{dp13}
\end{equation}
in the limit $k=0$. The analytical expression associated with (\ref{dp13})
is given by 
\begin{eqnarray}
&&\left[ \xi _{0}^{z}(t)\right] =\sqrt{L}g_{k}t+\mu
^{2}l^{-2}\sum\limits_{k=1}^{\infty }\int_{q}(-iq^{\alpha })C^{z\alpha
}(q)\int_{0}^{L}ds_{1}\int_{0}^{L}ds_{2}\int_{0}^{t}dt_{2}%
\int_{0}^{t_{2}}dt_{1}Q_{s_{2}0}Q_{s_{1}k}Q_{s_{2}k}  \nonumber \\
G_{k}^{0}(t_{2}-t_{1}) &&\exp (iq_{z}g_{k}(t_{1}-t_{2})+iq^{\prime
}(s_{2}-s_{1})(1-(s_{1}+s_{2})/2L)).  \label{dp14}
\end{eqnarray}
After performing in (\ref{dp14}) the integrations over times and arc lengths
we arrive at 
\begin{equation}
v_{c}(t)=\mu F/L(1-\sqrt{2\gamma }\frac{L^{3/2}}{l^{2}F^{5/2}}\int_{q}\frac{%
C^{zz}(q)}{\sqrt{\left| q_{z}\right| }}\int_{0}^{\sqrt{L\left| q_{z}\right|
\mu F/(2\gamma \mu )}}dx/(x^{4}+1)+...).  \label{dp15}
\end{equation}
The upper limit in the integration over $x$ can be put to infinity for large
driving force $F$, so that (\ref{dp15}) simplifies to 
\begin{equation}
v_{c}(t)=\mu F/L(1-c_{v}\frac{\sqrt{\gamma a}L^{3/2}}{l^{2}F^{5/2}}(\Delta
/a^{d})+...),  \label{dp16}
\end{equation}
with $c_{v}=\pi ^{2}/(\sqrt{2}\Gamma (3/4))$. Eq.(\ref{dp16}) gives the
first-order correction to the velocity, so that the correction term in (\ref
{dp16}) is expected to be small. This condition is fulfilled for large
driving force $F$. For finite $F$ and large $L$ the disorder drastically
affects the behaviour of the polymer chain. According to Eq.(\ref{dp16}),
the validity of the perturbation expansion for a given force $F$ is
restricted to the arc length of the polymer $L$, $L\leq F^{5/3}$.

Let us compare Eq.(\ref{dp15}) with the driven motion of an interface in
disordered media \cite{nstl92}. The expansion parameter of the perturbation
expansion (\ref{dp16}), $\sqrt{\gamma a}L^{3/2}/(l^{2}F^{5/2})(\Delta
/a^{d}) $, depends on $F$ as $F^{-5/2}$ instead of $F^{-3/2}$ as it is the
case for interfaces in $d=1$. The additional factor $1/F$ for the polymer
problem is a consequence of nontrivial integrations over $s_{1}$ and $s_{2}$%
, which are due to the absence of the factor $\delta ^{(d)}(x_{1}-x_{2})$ in
the disorder correlator (\ref{dp2}) and to the fact that the driving force
acts on the extremity of the polymer. The additional factor $1/F$ has the
consequence that the expansion parameter cannot be expressed in terms of the
force density $F/N$. \ The latter expresses the nontrivial carrying over of
the driving force $F$ to the monomers of the polymer chain.

To study the longitudinal size of the polymer we consider Eq.(\ref{dp13})
with $k\neq 0$. The analytical expression associated with the latter is 
\begin{eqnarray}
&&\left[ \xi _{p}^{z}(t)\right] =\frac{\sqrt{L}g_{p}}{(\gamma \mu p^{2}}+\mu
^{2}l^{-2}\sum\limits_{k=1}^{\infty }\int_{q}(-iq^{\alpha })C^{z\alpha
}(q)\int_{0}^{L}ds_{1}\int_{0}^{L}ds_{2}\int_{0}^{t}dt_{2}%
\int_{0}^{t_{2}}dt_{1}Q_{s_{2}p}Q_{s_{1}k}Q_{s_{2}k}  \nonumber \\
&&G_{p}^{0}(t-t_{1})G_{k}^{0}(t_{2}-t_{1})\exp
(iq_{z}g_{k}(t_{1}-t_{2})+iq^{\prime }(s_{2}-s_{1})(1-(s_{1}+s_{2})/2L)).
\label{dp17}
\end{eqnarray}
The average value $\left[ r^{z}(0)-r^{z}(L)\right] \equiv $ $\left[ \delta
r^{z}(L)\right] $ is obtained in terms of the normal coordinates as $2\sqrt{%
2/L}\sum_{k=1}^{\infty }\left[ \xi _{2k-1}^{z}\right] $. Performing the
integrations over the times and arc lengths results in 
\begin{equation}
\left[ \delta r^{z}(L)\right] =\frac{\sqrt{2}\mu FL}{4\mu \gamma }(1-c_{r}%
\frac{\sqrt{\gamma a}L^{3/2}}{l^{2}F^{5/2}}(\Delta /a^{d})+...)  \label{dp18}
\end{equation}
with $c_{r}=\pi ^{2}/(2\Gamma (3/4))$. Notice that the disorder weakens the
stretching of the polymer. This can be understood as follows. It is evident
that the effect of the disorder results in a decrease of the unperturbed
velocity $\mu F/L$. The decrease of the velocity due to disorder can be
interpreted as the unperturbed motion due to a smaller force, which in
virtue is accompanied by a less stretching of the polymer. Thus, this
argument supports the non evident result of Eq.(\ref{dp18}) that the
stretching of the polymer in the longitudinal direction will be diminished
due to disorder.

To conclude, we have considered the drift of a polymer chain in disordered
media under the action of a constant force applied to one of the extremities
of the polymer by neglecting the thermal effects. The dynamics of the
polymer is governed by the interplay between the motion and the quenched
random forces. One of the significant effects of this interplay is the
generation of transversal fluctuations of the polymer chain. The latter take
also place for a Brownian particle. We have also found that disorder results
in a weakening of the elongation of the polymer in the longitudinal
direction, which is caused by the friction of the polymer with the
surrounding medium. Due to the specific dependence of the quantities under
investigation on the driving force $F$ and the length of the polymer $L$,
the fluctuations induced by the drift of the polymer may compete with
thermal fluctuations. An interesting question is the experimental
manifestation of these fluctuations. However, the first-order corrections
studied in the present Letter do not allow us to make predictions on the
total effect of the disorder on the quantities under consideration. The only
conclusion we may drawn on this matter is that for finite driving force the
effect of disorder increases with the polymer length. The transversal
fluctuations studied in the present Letter may be also of relevance for the
related problems of the drift of a directed polymer in disordered media,
crack propagation in disordered media etc. To our knowledge, the transversal
fluctuations in these problems have not been studied so far.

\acknowledgments A support from the Deutsche Forschungsgemeinschaft (SFB
418) is gratefully acknowledged. Useful comments and suggestions of the
Referee are also acknowledged.

\end{document}